\documentclass[10pt,journal,comsoc]{IEEEtran}
\usepackage{url}

\usepackage[noadjust]{cite}
\usepackage{graphicx}
\usepackage{multirow}

\usepackage{subcaption}
\usepackage{xspace}

\usepackage{makecell}

\usepackage{rotating}
\usepackage{lscape}
\usepackage{balance}
\usepackage[utf8]{inputenc}

\usepackage{lipsum}
\usepackage{array, tabularx, multirow, booktabs, diagbox}
\usepackage{adjustbox}
\usepackage{amsmath,amssymb,amsfonts}

\begin{document}
\title{Dispersed Federated Learning: Vision, Taxonomy, and Future Directions}

\author{Latif~U.~Khan,~Walid~Saad,~\IEEEmembership{~Fellow,~IEEE},~Zhu~Han,~\IEEEmembership{~Fellow,~IEEE},~and~Choong~Seon~Hong,~\IEEEmembership{Senior~Member,~IEEE}

\IEEEcompsocitemizethanks{
\IEEEcompsocthanksitem L.~U.~Khan~and~C.~S.~Hong are with the Department of Computer Science \& Engineering, Kyung Hee University, Yongin-si 17104, South Korea.

\IEEEcompsocthanksitem Walid Saad is with the  Wireless@VT, Bradley Department of Electrical and Computer Engineering, Virginia Tech, Blacksburg, VA 24061 USA

\IEEEcompsocthanksitem Zhu Han is with the Electrical and Computer Engineering Department, University of Houston, Houston, TX 77004 USA, and also with the Computer Science Department, University of Houston, Houston, TX 77004 USA, and the Department of Computer Science and Engineering, Kyung Hee University, South Korea.

}}

\markboth{}{}%

\maketitle






\begin{abstract} 
The ongoing deployments of the Internet of Things (IoT)-based smart applications are spurring the adoption of machine learning as a key technology enabler. To overcome the privacy and overhead challenges of centralized machine learning, there has been a significant recent interest in the concept of federated learning. Federated learning offers on-device machine learning without the need to transfer end-devices data to a third party location. However, federated learning has robustness concerns because it might stop working due to a failure of the aggregation server (e.g., due to a malicious attack or physical defect). Furthermore, federated learning over IoT networks requires a significant amount of communication resources for training. To cope with these issues, we propose a novel framework of dispersed federated learning (DFL) that is based on the true decentralization. We opine that DFL will serve as a practical implementation of federated learning for various IoT-based smart applications such as smart industries and intelligent transportation systems. First, the fundamentals of the DFL are presented. Second, a taxonomy is devised with a qualitative analysis of various DFL schemes. Third, a DFL framework for IoT networks is proposed with a matching theory-based solution. Finally, an outlook on future research directions is presented.                              
\end{abstract}



\section{Introduction}
\setlength{\parindent}{0.7cm}Recent years have revealed a significant rise in the number of Internet of Things (IoT) devices to enable various applications, such as smart health-care, augmented reality, industry 4.0, autonomous driving cars, among others. These applications use emerging communication and computing technologies along with machine learning to offer smart services. In order to deploy machine learning in large-scale, heterogeneous systems such as the IoT, it is necessary to preserve the privacy of the data and reduce the communication overhead. As a result, centralized machine learning techniques may not be suitable. Instead, \textit{federated learning (FL)} ~\cite{mcmahan2016communication} is a distributed machine learning solution that can be amenable to deployment in an IoT. Although FL enables on-device machine learning, it faces a few challenges.
\begin{itemize}
    \item Traditional FL based on a centralized aggregation server might suffer from a malicious user attack or failure due to a physical damage, which significantly degrades the performance of FL. The aggregation server can be attacked by (a) an outsider that is not participating the learning process or (b) one of end-devices participating in the learning process. 
    \item A malicious aggregation server can infer the end-devices sensitive information from their learning model parameters \cite{lim2020federated}. Therefore, there is a need to address the privacy leakage challenge of FL. 
    \item FL requires a significant amount of communication resources for the iterative exchange of learning model parameters between massive number of devices and the aggregation server.
\end{itemize}\par
To truly benefit from the deployment of FL in IoT networks, we must resolve the aforementioned challenges. To address the privacy concerns of FL, a user-level differential privacy protection obtained by adding noise to local learning models was proposed in \cite{mcmahan2017learning}. Although the differential privacy scheme in \cite{mcmahan2017learning} can offer privacy preservation, it may suffer from prolonging the global FL model convergence time. Other works in \cite{koda2020differentially} and \cite{liu2020privacy} proposed privacy preserving schemes based on over-the-air-computation. The works in \cite{koda2020differentially} and \cite{liu2020privacy} can preserve the privacy but at the cost of additional design complexity. On the other hand, a number of recent works (e.g., \cite{chen2019joint, Fedlearning_edge_1,abad2020hierarchical}) considered resource optimization in FL. However, these works (i.e., \cite{chen2019joint, Fedlearning_edge_1}, and \cite{abad2020hierarchical}) did not discuss FL privacy issues. In \cite{chen2020wireless}, a novel framework, called collaborative FL (CFL) was proposed to enable the participation of devices with insufficient communication resources in the learning process for performance enhancement. In CFL, the devices send their local learning models to nearby devices with sufficient communication resources for local aggregation. Subsequently, the receiving devices send the locally aggregated models to the base station (BS) for global aggregation. The BS after performing global aggregation sends back the global model updates to the end-devices for local models update. On the other hand, hierarchical FL proposed in \cite{abad2020hierarchical} has multiple iterative sub-global aggregations at various small cell base stations (SBSs) and global aggregation at macrocell BS. All of three works, such as traditional FL \cite{mcmahan2016communication}, hierarchical FL \cite{abad2020hierarchical} and CFL \cite{chen2020wireless} suffer from a robustness issue. In contrast, the main contribution of this paper is a novel framework, dubbed \textit{dispersed FL (DFL)} that offers learning of a global FL model in a fully distributed manner. The DFL framework uses a distributed fashion of learning to jointly offer efficient communication resources reuse, robustness, and enhanced privacy for FL. In our proposed DFL, first, sub-global models are computed within different groups consisting of closely located end-devices. The sub-global models are then aggregated to yield a global model. Aggregation, as discussed in Section \ref{taxonomy}, can be either centralized or distributed depending on the type of DFL. The distributed aggregations of sub-global models will result in a robust operation in contrast to traditional FL that is based on a centralized aggregation. Finally, the global model updates are sent back to the end-devices. The two-stage aggregation of learning models in DFL can offer a better privacy protection compared to the traditional FL. Inferring end-devices sensitive information from sub-global models (e.g., at global server) is very difficult compared to inferring of information at sub-global model computation server using local learning model updates \cite{nasr2019comprehensive}. Thus, we can say that DFL can offer a better privacy preservation. Furthermore, one can reuse the communication resources occupied by other cellular users within small groups used for sub-global model computation to offer efficient communication resource usage. The summary of our contributions are as follows:\par
\begin{itemize}
    \item We present a framework of DFL and proposed a taxonomy using the approach of aggregating of sub-global models to yield a global model, as a parameter.
    \item We propose a cost function for DFL that captures loss in global model accuracy due to packet error rate and local learning model accuracy. To minimize the cost of DFL, we propose an iterative scheme that performs joint association and resource allocation. For association, we use a one-sided one-to-many matching game, whereas a one-sided one-to-one matching game is used for resource allocation. 
    \item To validate our proposed iterative-matching-game-enabled solution, we provide numerical results. We show the fast convergence of the proposed scheme for a fewer number of global DFL rounds. Furthermore, we evaluate the performance of DFL using the MNIST dataset for image classification tasks, which shows promising results.
    \item Finally, we conclude the paper and provide an outlook on future research directions. 
    
\end{itemize}

\begin{figure*}[!t]
	\centering
	\captionsetup{justification=centering}
	\includegraphics[width=12cm, height=12cm]{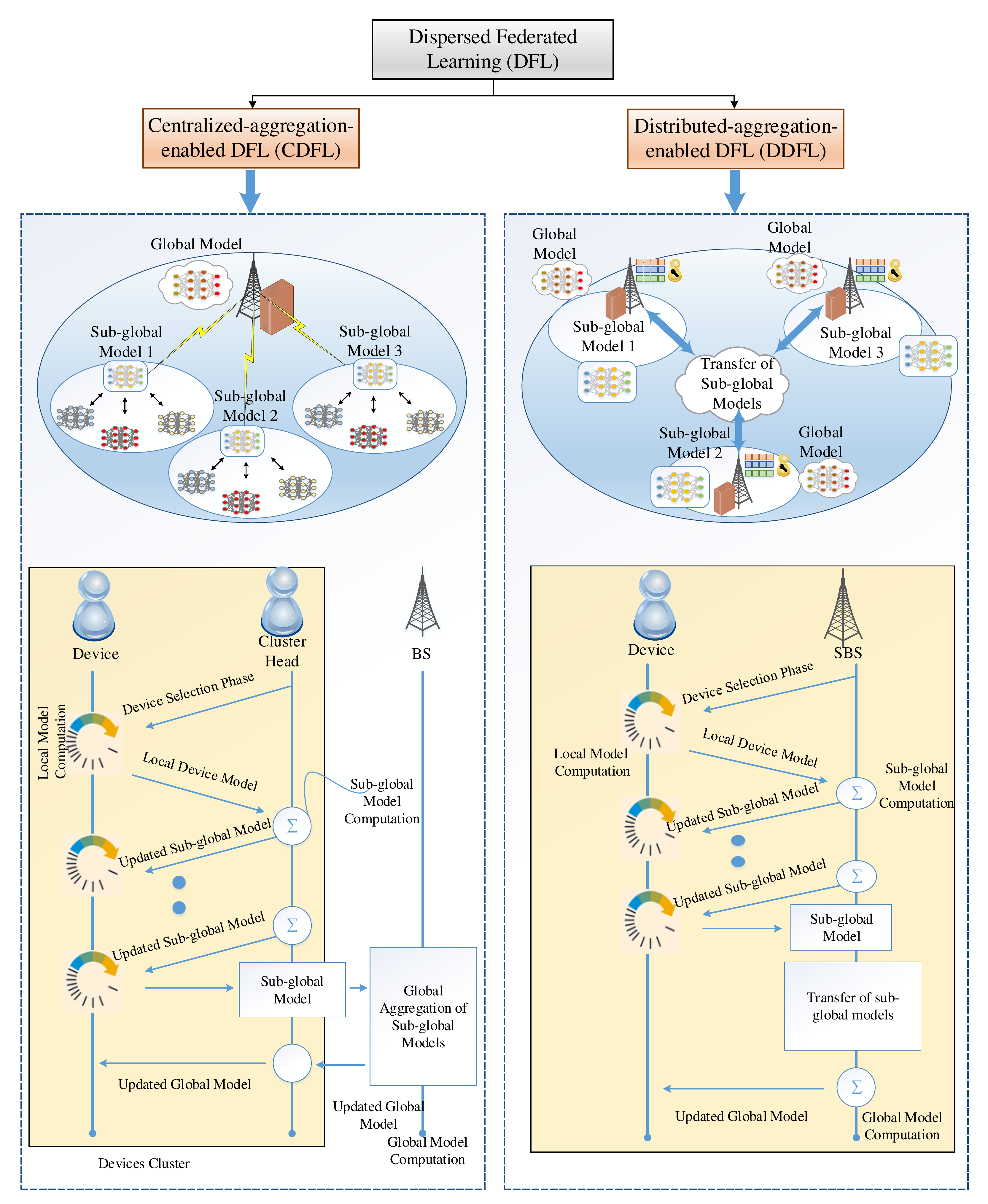}
	\caption{Taxonomy of DFL}
	\label{fig:taxonomy}
\end{figure*}

\section{DFL: Fundamentals and Taxonomy}
\label{taxonomy}
\subsection{Fundamentals}
\setlength{\parindent}{0.7cm}In DFL, a sub-global model is first iteratively computed for different groups similar to traditional FL. Next, the sub-global model updates are aggregated to yield to a global model. The aggregation of the sub-global models can be performed either in a centralized similar to hierarchical FL or a distributed way. Finally, the global model updates are sent back to all the devices involved in learning as shown in Fig.~\ref{fig:taxonomy}. Here, learning takes place in an iterative manner until convergence of the global FL model to a desirable value. When using a centralized global aggregation server, DFL can be considered to be hierarchical FL with sub-global aggregation at small cell base stations (SBSs) and global at a macrocell base station (MBS) \cite{abad2020hierarchical}. The work in \cite{abad2020hierarchical} considered a hierarchical FL scheme that performs global aggregation of sub-global models at the MBS. Hierarchical FL generally follows the leader-follower concept, where cloud/MBS acts as a leader and leads the learning process \cite{abad2020hierarchical,zhao2020federated,hosseinalipour2020federated}. The leader will announce to the followers about participation by broadcasting the messages. After receiving the acknowledgment from the followers, the leader will decide on the participation of devices in the learning process along with initialization parameters (e.g., fixed sub-global iterations for all groups). In contrast, DFL will have multiple groups without following the single leader with multiple followers trend. Within every group, the sub-global aggregator can act as a leader and control the learning within that group. To learn a global model, different groups can coordinate with each other about the model initialization and sharing of sub-global model updates. DFL provides more freedom for performing sub-global model learning by using different selection criteria of end-devices with other parameters (e.g., sub-global iterations). Additionally, the trust verification criteria for participation of devices in learning can also be distinct for every group. On the other hand, the sub-global model iterations can also be different depending on the computational and communication resource constraints of end-devices within groups used for sub-global model computation. Performing more sub-global iterations yields better performance in terms of learning model accuracy. Such a fashion of variable sub-global iterations can offer an additional advantage of fairness. For fixed sub-global model iterations, some of the groups might not perform well in terms of accuracy due to data and system heterogeneity. The sub-global model computing groups with high accuracy will affect the global model more compared to the poor-performing sub-global model computing groups. Therefore, groups with high computing resources can perform more sub-global iterations to improve their performance to enable fairness-based DFL. Overall, DFL can better scale with the number of devices.\par
For a fixed global FL model accuracy, there is a tradeoff between the number of sub-global iterations and the number of global rounds. An increase in sub-global model iterations requires few global rounds to train the FL model, and vice versa. The selection of sub-global model iterations and global rounds strongly depend on the application and settings used. For instance, consider an intelligent transportation setting in which the sub-global model computation takes place at the level of an autonomous vehicle and the global model aggregation at the core network. For such scenarios, it is preferable to use more sub-global iterations than global iterations due to the high mobility of vehicles and the potentially high number of handovers between the roadside units (RSUs). On the other hand, consider the scenario where different smart industries seek to train a DFL model for a certain application. First, we can train a sub-global model within every industry. Next, the sub-global models are shared between the industries. Finally, global aggregation takes place within every industry and global model updates are sent back to end-devices. In such a scenario of smart industries, we can use a few sub-global iterations and more global rounds compared to the later scenario of autonomous vehicles.\par
Now we discuss the advantages of DFL. First, DFL provides better privacy preservation than the traditional FL. In DFL, there are two aggregations, such as sub-global aggregation and global aggregation. A sub-global aggregator can infer the end-devices sensitive information using their local updates \cite{lim2020federated}, whereas it is very difficult for a global aggregation server to infer the end-devices sensitive information from the sub-global model updates. On the other hand, in traditional FL there is only one aggregation that can easily infer the end-devices information from the learning model updates. Therefore, we can say that DFL can preserve privacy better than the traditional FL. Furthermore, DFL can offer more robust operation than the FL. In contrast to FL, DFL performs multiple sub-global aggregations. These multiple aggregations are followed by global aggregation that can be either centralized or distributed. A distributed aggregation of sub-global model updates will offer the highest robustness as given in Table~\ref{tab:comparison}. Other than robustness and privacy preservation, the DFL trains multiple sub-global models in distributed groups. Therefore, one can efficiently reuse the channel resources that are already in use by other cellular users by keeping the interference level below the maximum allowed limit \cite{kazmi2017mode}. Collectively, we can say that DFL can provide robust operation, better privacy preservation, and efficient reuse of communication resources than the FL.


\subsection{Taxonomy}
\setlength{\parindent}{0.7cm}We can categorize DFL into two main types (shown in Fig.~\ref{fig:taxonomy}) depending on the fashion of global model aggregation: CDFL and DDFL, as discussed next.\par

\begin{table*}

\caption {Comparison of FL and various DFL schemes.} 
\label{tab:comparison} 
  \centering
  \begin{tabular}{p{2cm}p{5.6cm}p{1.5cm}p{1.5cm}p{1.5cm}p{1.5cm}p{1.5cm}}
    \toprule
\textbf{Parameter }&\textbf{Description}& \textbf{Edge-based CDFL}  & \textbf{Cloud-based CDFL} & \textbf{Edge-based DDFL} & \textbf{Blockchain-based DDFL}&\textbf{Traditional cloud-based FL}\\ \toprule

\textbf{Global model computation delay }& This is related to the delay in computing global model during for one global round. & Lowest & Low  &  High & Highest & High\\ \midrule 


\textbf{Robustness }& It refers to  a successful operation of DFL in case of edge server/cloud server/miners physical damage. Additionally, robustness is also related to the interruption of DFL due to malicious attack on the aggregation server. & Low & Low  & High  & Highest & Lowest\\ \midrule   

\textbf{Communication resources usage }& This refers to the communication resources used for computing global DFL model.  & Low & High & High & Highest & Higher\\ \midrule

\textbf{Implementation complexity}& This is the measure of the complexity in the computation of a global DFL model. & Lowest & Low  & High  & Highest & Low\\ 


\bottomrule

\end{tabular}
\end{table*}

\subsubsection{Centralized-Aggregation-Enabled DFL}
\setlength{\parindent}{0.7cm}In CDFL, a global FL model is obtained by aggregation of the sub-global models at a centralized server. In contrast to CFL \cite{chen2020wireless}, CDFL iteratively computes sub-global models for all groups of closely located devices prior to global aggregation. Meanwhile, in CFL, few single iteration local aggregations at a few devices take place prior to global aggregation at the BS. CDFL can be either edge-based or cloud-based depending on the context of the machine learning model. Edge-based CDFL is more suitable for the local context-aware, specialized machine learning model for devices located within a small geographical region. For instance, we can train a local context-aware, specialized machine learning model (e.g., a smart keyboard search suggestion for a regional language) using edge-based CDFL. To do so, we can consider an edge-based SBS and a device-to-device (D2D) communication network reusing the already occupied frequency bands by other cellular users. Multiple clusters with their own cluster heads can be formed using some criteria (e.g., throughput enhancement). First, sub-global models can be trained in a fashion similar to traditional FL for different clusters with the cluster head acting for sub-global model aggregation. Second, sub-global models are transferred to the edge servers for global model aggregation by cluster heads. Finally, the global model updates are transmitted back to the cluster heads which disseminate them to all the devices. On the other hand, cloud-based DFL can train global context-aware, generalized machine learning models for devices located within several geographically distributed regions. The process of cloud-based DFL is similar to edge-based DFL, but with the aggregation of sub-global models taking place at the remote cloud. Although cloud-based DFL enable more generalized learning of a model, it suffers from the high-latency issue. Furthermore, comparative analysis with DFL schemes are also provided in Table~\ref{tab:comparison}.\par      

\subsubsection{Distributed-Aggregation-Enabled DFL}
\setlength{\parindent}{0.7cm}DDFL aggregation of the sub-global model updates is performed at distributed nodes in contrast to CDFL. CDFL is based on centralized aggregation of sub-global models which can be suffered from centralized server crash or a malicious attack. DDFL overcomes these issues by using distributed nodes-based aggregation. However, it suffers from high latency and extra communication resources for transfer of sub-global model updates among distributed nodes, as given in Table~\ref{tab:comparison}. To implement DDFL, there can be two possible ways depending on the way in which the sub-global model updates are exchanged. The sharing of sub-global model updates can be either using blockchain-based miners or direct exchange between edge servers. Blockchain-based DFL can offer the most trustful transfer of sub-global model updates. However, it faces the inherent challenge of high latency associated with a blockchain consensus algorithm (e.g., proof of work) that is not desirable in FL model computation \cite{lee2020letters}. Therefore, we must design novel consensus algorithms with low latency for blockchain-based DFL.  \par


\section{DDFL Framework for IoT Systems}
\label{framework}
For an IoT system-enabled by FL using a massive number of end-devices, we need a significant amount of communication resources for training. To enable communication resource-efficient and robust FL for such an IoT network, we can either use CDFL or DDFL. In contrast to CDFL, DDFL performs sub-global model aggregations at multiple locations and thus offers a more robust operation.
\subsection{System Model}
\setlength{\parindent}{0.7cm} Consider an IoT network which consists of IoT devices and edge server-enabled SBSs. The IoT devices, each having its own local datasets seek to train FL model. The IoT devices show significant heterogeneity in terms of computational power (CPU-cycles/sec), local dataset size, and data distribution. To account for IoT devices heterogeneity, we use the notion of relative local accuracy. The lower values of relative accuracy reflect better local learning model accuracy, and thus less global communication rounds are needed to achieve a desirable global FL accuracy. To enable edge-based DDFL in IoT network, first, a sub-global model is computed for SBS in an iterative fashion via interaction with its associated IoT devices. After the computation of the sub-global model updates by all the SBSs, the sub-global model updates are transferred between all the SBSs via fast backhaul links. Finally, all the SBSs perform global model aggregation and sent the global model updates to their associated IoT devices. The process for edge-based DDFL uses wireless channel resources. The packet error rate due to channel uncertainties might cause severe degradation effect on the performance of the DDFL process \cite{chen2019joint}. We define a cost function that jointly accounts for relative local accuracy and the degradation effect due to the packet error rate on the performance of DDFL. The cost function is given by taking the product of the relative accuracy term (i.e., $1$+relative accuracy) and packet error rate term (i.e., $1$-exponential function with an exponent of $(-1/SINR)$ multiplied by a waterfall threshold). For high relative local accuracy, the local learning model accuracy will be low and vice versa. Therefore, lower values of relative local accuracy are desirable. The packet error rate strongly depends on the device's signal-to-interference-plus-noise ratio (SINR). Therefore, to enable edge-based DDFL for IoT networks while minimizing the cost function, there is a need to address the two following challenges.\par
\begin{itemize}
\item An IoT network is characterized by a dynamic topology due to the free mobility of IoT devices. Furthermore, the seamless connectivity of IoT devices with the SBSs is desirable during the exchange of learning model parameters. Considering the aforementioned challenges, we must appropriately associate IoT devices with corresponding SBSs so as to optimize FL performance by increasing the overall throughput that consequently minimizes the cost.\par  

\item To effectively use the limited available communication resources, we can reuse the already occupied resource blocks by cellular users for IoT-networks. Therefore, we must propose an effective resource allocation strategy for FL-based IoT networks.\par  
\end{itemize}

\subsection{Iterative-Matching-Based Solution}
\setlength{\parindent}{0.7cm}In this subsection, we present an iterative scheme for cost optimization of proposed edge-based DDFL. First, we consider several constraints for our DDFL framework:\par
\begin{itemize}
    \item \textit{Constraint 1:} The size of the packet used to transfer the local learning model of end-devices to the aggregation server strictly depends on the architecture of the local learning model. The architecture of local learning model is dependent on the application and dataset. Therefore, we assume that every device can be allocated a maximum of one resource block.
    \item \textit{Constraint 2:} Every resource block must not be allocated to more than one device.
    \item \textit{Constraint 3:} The total number of resource blocks allocated to all devices must not be greater than the maximum available resource blocks.
    \item \textit{Constraint 4:} Every device be associated to a maximum of one SBS.
    \item \textit{Constraint 5:} Every SBS can be associated with maximum number of devices determined by a threshold (fixed in this paper for all SBSs).
\end{itemize}\par
\setlength{\parindent}{0.7cm}We consider a cost function which jointly considers devices relative local accuracy and effect of packet error rate on global DDFL model accuracy. The cost function jointly depends on both resource allocation and devices association. Due to the NP-hard nature of the joint optimization problem for optimizing both resource allocation and device association, we decompose the main problem into two sub-problems: resource allocation and association sub-problems. Our proposed algorithm solves one problem (i.e., resource allocation) by fixing the other problem (i.e., association) in an iterative manner until the convergence.\par 

\subsubsection{One-to-One Matching-Based Resource Allocation}
\setlength{\parindent}{0.7cm} To minimize the DDFL cost (i.e., improving FL accuracy) for IoT networks by improving SINR, we can use a low-complexity matching theory-based wireless resource allocation scheme. Our resource allocation problem is similar to house allocation problem that can be represented by a tuple $(\mathcal{A},\mathcal{H},\mathcal{P})$, where $\mathcal{A}$, $\mathcal{H}$, and $\mathcal{P}$ represent the agents set, houses set, and preference profile of agents over houses, respectively \cite{zhou1990conjecture}. Similar to the house allocation problem, we can define a preference profile $\mathcal{P}_{r}$ for a set of resource blocks over devices. To allocate resource blocks to devices, we use a one-sided one-to-one matching game. All the resource blocks rank devices based on the values of the cost function. For a particular resource block, the device with low cost is preferred over the car with a high cost. Using preference profile $\mathcal{P}_{r}$, all the resource blocks are allocated to devices iteratively until no blocking pair is left, which shows stable matching \cite{han2012game}.\par

\subsubsection{One-to-Many Matching-Based Devices Association}
\setlength{\parindent}{0.7cm}For a fixed resource block allocation, the DDFL cost strongly depends on the association of devices with SBSs. Therefore, we must associate the devices with SBSs so as to minimize the  DDFL cost by increasing the throughput. Moreover, it will decrease DDFL convergence time. We can associate devices to SBSs using a brute force algorithm, it suffers from high computational complexity. In contrast to the house allocation problem, our device association problem involves the assignment of a single SBS to multiple devices. Therefore, we use one-sided one-to-many matching for the association of SBSs to devices. First, we define a preference profile $\mathcal{P}_{a}$ for RSUs. The preference profile $\mathcal{P}_{a}$ ranks all the devices based on the values of the cost function. Similar to $\mathcal{P}_{r}$, the device with a high cost is given less preference than the device with a low cost. Once the preference profile is obtained for all devices, the association is performed in an iterative manner until no blocking par is left to achieve stable matching \cite{han2012game}.\par

\begin{figure*}[t!]
	\centering
	\begin{subfigure}[b]{0.3\textwidth}
		\includegraphics[width=2in]{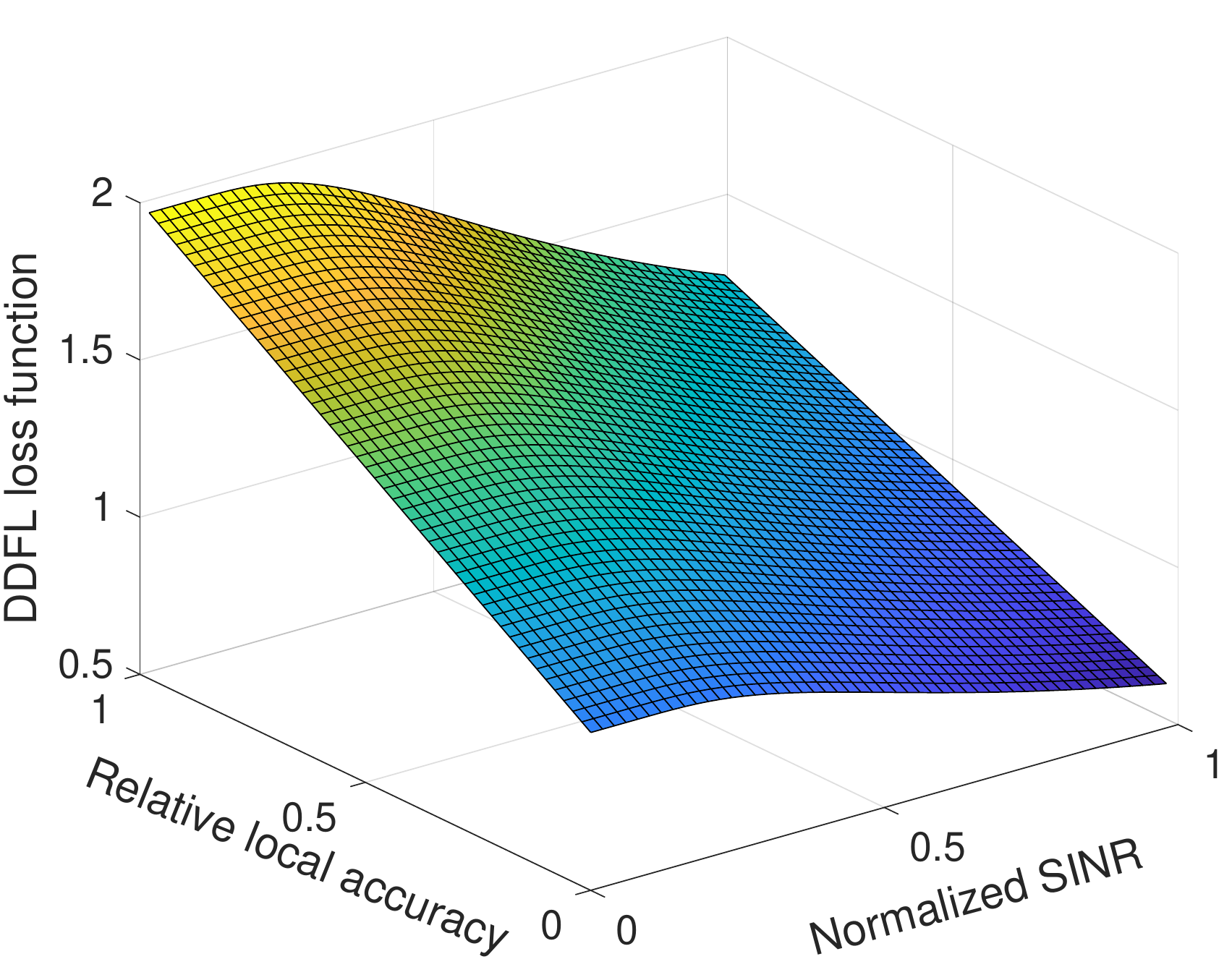}
		\caption{}
	\end{subfigure}%
    	\hfill
	\begin{subfigure}[b]{0.3\textwidth}
		\includegraphics[width=2in]{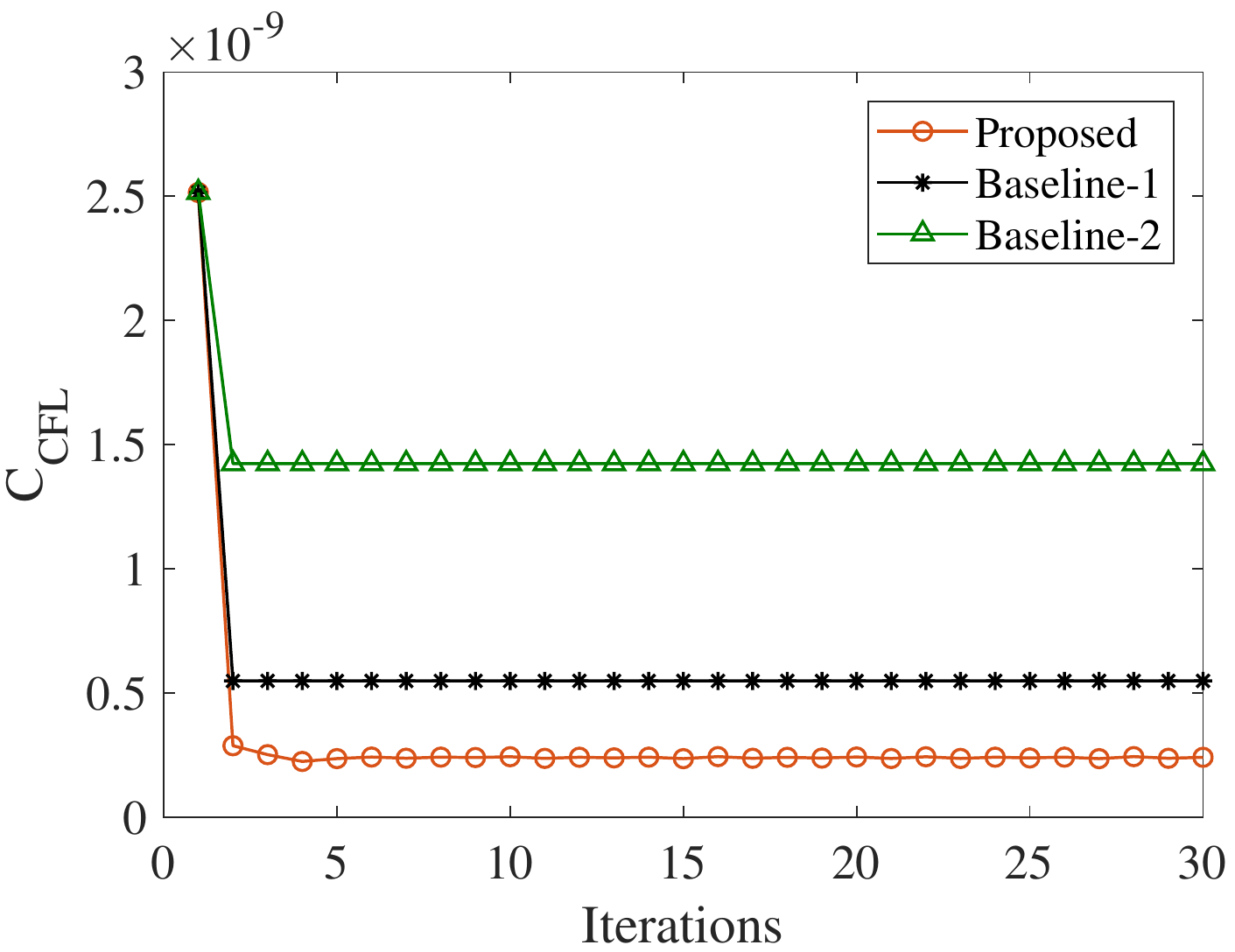}
		\caption{}	
	\end{subfigure}
		\hfill
	\begin{subfigure}[b]{0.3\textwidth}
		\includegraphics[width=1.9in]{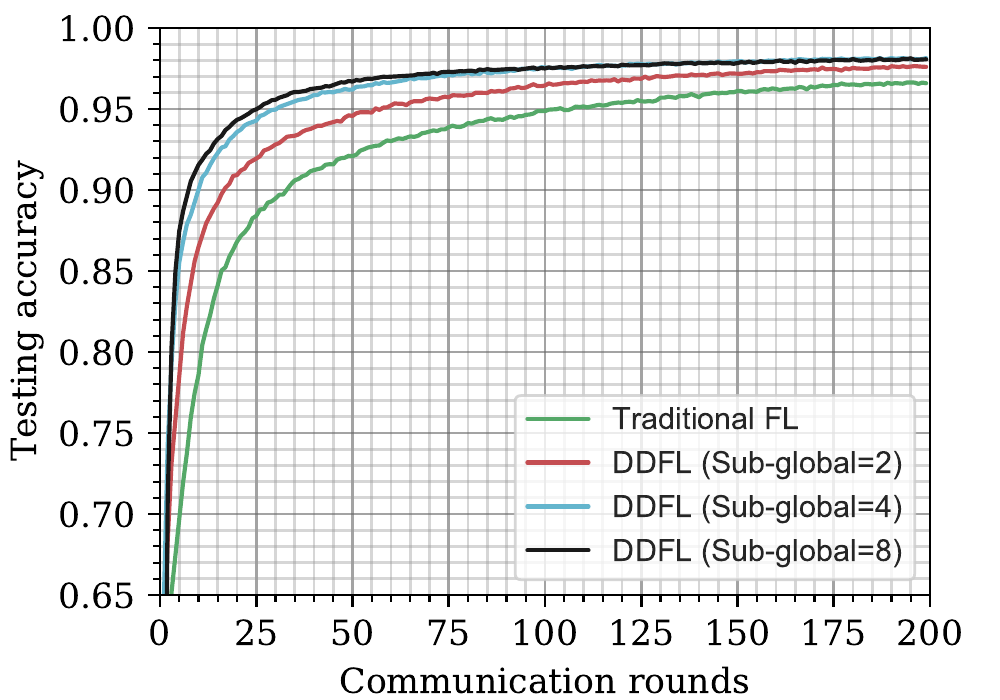}
		\caption{}	
	\end{subfigure}
	\centering
	\caption{(a) Variations in DDFL cost for SINR and relative local accuracy,(b) Proposed DDFL cost vs. iterations, (c) Accuracy vs. communication rounds for non-IID data \cite{mcmahan2016communication} and $T=8$.}
	\label{fig:results}
\end{figure*}

\subsection{Performance Evaluation}
\setlength{\parindent}{0.7cm}We present numerical results to evaluate the performance of the proposed framework for IoT networks. Moreover, we evaluate the performance of the proposed DFFL framework for an image classification task using the MNIST dataset \cite{mcmahan2016communication}. Our simulation scenario consists of $54$ devices and $6$ SBSs which are deployed in an area of $1000\times 1000~m^2$. The position of the SBS is taken fixed whereas devices are positioned randomly. All the values are generated using an average of $500$ different runs. Other parameters such as carrier frequency and number of sub-carriers per resource block are taken $2$GHz and $12$, respectively. Furthermore, the free space path loss model is considered in our work \cite{kazmi2017mode}. The position of the SBS is taken fixed whereas devices are deployed randomly. All statistical results are generated using an average of 500 different runs. We compare the performance of our proposed iterative-matching-based scheme with two baselines schemes. One-to-many matching-based association and random resource allocation are used for baseline-1, whereas random association and one-to-one matching-based resource allocation are used for baseline-2. We use the term "iterations" to refer to one-time execution of both one-to-one matching-based resource allocation and one-to-many matching-based association. Furthermore, the keyword "communication round" denotes global iteration for DDFL. \par
\setlength{\parindent}{0.7cm}Fig.~\ref{fig:results}a shows the variations in DDFL cost for signal-to-interference-plus-noise-ratio (SINR) and relative local accuracy. For a fixed relative local accuracy, the DDFL cost decreases exponentially with an increase in SINR and vice versa. For lower values of SINR, the decrease in DFL cost is very less, but for higher values of SINR, the cost decreases significantly. On the other hand, a decrease in relative local accuracy shows a decrease in DDFL cost. The DDFL cost shows lower values for higher and lower values of SNIR and relative local accuracy, respectively. This shows that the cost of training DDFL global model is low for devices having high local learning model accuracy (low relative local accuracy) and high throughput and vice versa. Fig.~\ref{fig:results}b shows superior performance of our proposed DDFL scheme over baseline-1, and baseline-2 schemes. Moreover, all three schemes showed fast convergence for a lower number of iterations. Baseline-1 has better performance than baseline-2, which shows that DDFL cost depends more on devices association than resource block allocation.\par
Finally, Fig.~\ref{fig:results}c evaluates the performance of DDFL using the MNIST dataset for image classification tasks using a convolutional neural network at the local devices. We consider $T$ as the product of local iterations and sub-global iterations for a single global round. For analysis, we use Non-IID distribution of data that is based on sorting and dividing the entire MNIST dataset into $200$ shards of $300$ images each and assigns $2$ shards to every device \cite{mcmahan2016communication}. FL here refers to traditional cloud-based FL using an equal number of end-devices and local iterations (i.e., 8) at every device, as DDFL. From Fig.~\ref{fig:results}c, we can clearly observe that DDFL outperforms traditional FL at all sub-global iterations. Fig.~\ref{fig:results}c shows that DDFL exhibits a convergence time that is much faster than traditional FL when using different numbers of sub-global iterations. For a fixed $T$, faster convergence is observed for higher number of sub-global iterations, and thus requiring few global rounds to reach a certain accuracy level. From the above discussion, we observe a tradeoff between the number of sub-global iterations and the number of global communication rounds for achieving a certain global FL accuracy. \par          





\section{Conclusions and Future Directions}
\setlength{\parindent}{0.7cm}In this paper, we have presented a novel idea of DFL. A taxonomy is devised based on the fashion of global model aggregation for DFL. We have found that DFL can be adopted widely to achieve resource-efficient and privacy-aware implementation of FL for several IoT scenarios. Furthermore, DFL offers a tradeoff between convergence rate and the number of sub-global iterations. Finally, we present several future research directions regarding DFL.\par   

\subsection{Data Heterogeneity-Aware DFL}
\label{DHA}
\setlength{\parindent}{0.7cm}How do we enable FL for a massive number of devices with heterogeneous datasets of non-IID nature? In contrast to FedAvg, FedProx was developed to account for data heterogeneity in the training of a global FL model. FedProx is based on the addition of a weighted proximal term to end-device loss function to handle data heterogeneity. However, choosing the weight of the proximal term might be challenging. Furthermore, it is unclear whether FedProx can provably improve the convergence rate \cite{kairouz2019advances}. To cope with these challenges, we can propose novel heterogeneity-aware DFL for devices with heterogeneous datasets. A set of devices can be divided into different clusters based on datasets homogeneity, with their own cluster heads acting for sub-global aggregation. The sub-global models are obtained in an iterative manner via the exchange of learning model parameters between the end-devices and their cluster heads. Then, the sub-global models are aggregated to yield a global model which is then sent back to all the cluster heads. Finally, the cluster heads disseminate the global model updates to all the devices for updating their local learning models. This process of global model computation takes place iteratively until convergence. It must be noted that homogeneous clustering-based DFL will converge faster than the FL because of two reasons. DFL converges faster than the FL as revealed by Section~\ref{framework}. Moreover, learning of a sub-global model for a homogeneous set of devices will minimize the convergence time \cite{mcmahan2016communication,lim2020federated}. \par


\subsection{Enhanced Distributed Privacy-Aware DFL}
How do we enable DFL with enhanced privacy preservation? Although the works in \cite{mcmahan2017learning, koda2020differentially, liu2020privacy} can be used to enable privacy preservation in FL, these algorithms may not be able to perform well for our proposed DFL framework. Different than FL, two kinds of weight divergences will occur in DFL: one between the end-devices and sub-global servers, and the other between sub-global servers and global aggregation server. In contrast to global aggregation server in DFL, the server used for sub-global aggregation can infer the end-devices sensitive information from their local learning models. Therefore, new effective privacy-preserving schemes within groups used for sub-global model computation must be designed. One way is to add artificial noise to local learning models before sending for sub-global aggregation. Other way can be use of wireless channel noise as a privacy preserving scheme for DFL \cite{koda2020differentially, liu2020privacy}.



\subsection{Mobility-Aware DFL}
How do we enable efficient DFL for mobile nodes? An end-device involved in DFL process requires seamless connectivity with the sub-global aggregation server during sub-global model computation. Therefore, it is necessary to propose novel protocols that accounts for end-devices mobility. Several ways can be used for mobility management such as deep learning-enabled mobility prediction, hidden Markov model, Bayesian network, and data mining. Deep learning models (e.g., Long Short Term Memory) can be used to learn devices mobility pattern using their previous traces. Hidden Markov model can also be used for devices mobility pattern because devices' mobility posses Markov property. Additionally, Bayesian network uses directed acyclic graph for mobility management of devices.

\bibliographystyle{IEEEtran}
\bibliography{Database}

\begin{IEEEbiography}[{\includegraphics[width=1in,height=1.25in,clip,keepaspectratio]{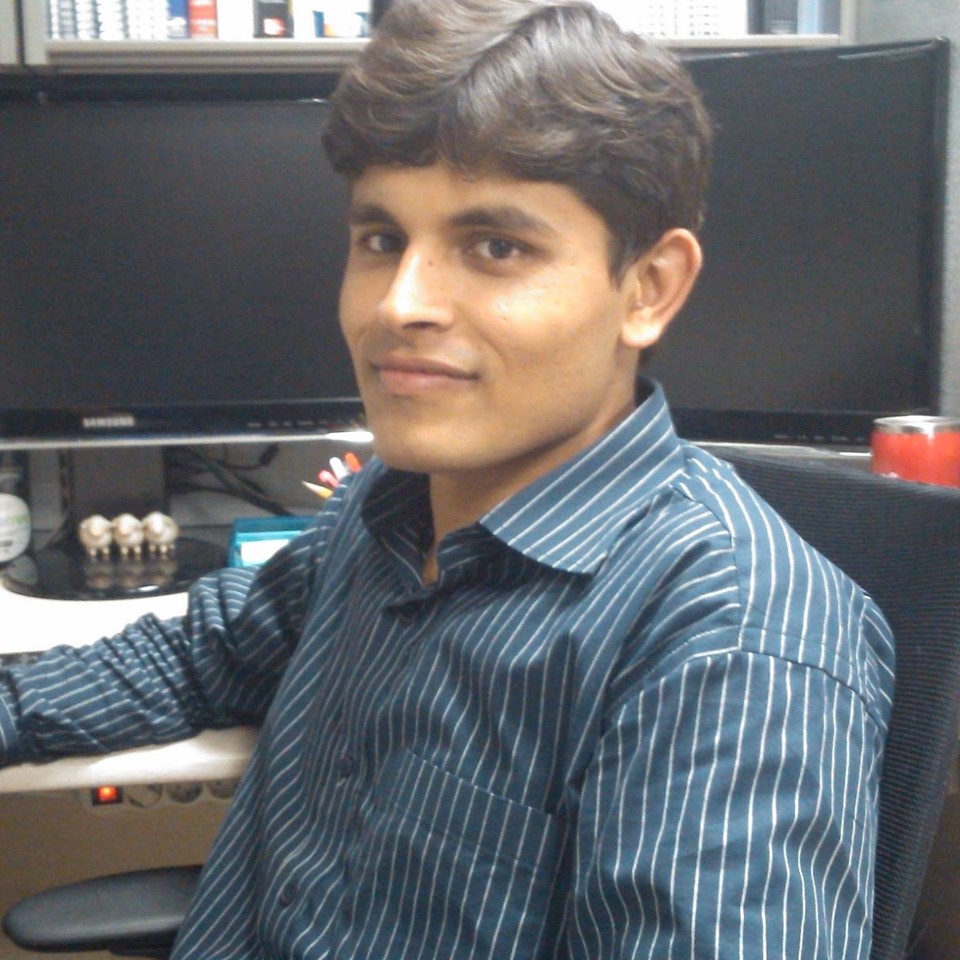}}]{Latif U. Khan} is currently pursuing his Ph.D. degree in Computer Engineering at Kyung Hee University (KHU), South Korea. He received his MS (Electrical Engineering) degree with distinction from University of Engineering and Technology, Peshawar, Pakistan in 2017. His research interests include analytical techniques of optimization and game theory to edge computing and end-to-end network slicing. 
\end{IEEEbiography}

\begin{IEEEbiography}[{\includegraphics[width=1in,height=1.25in,clip,keepaspectratio]{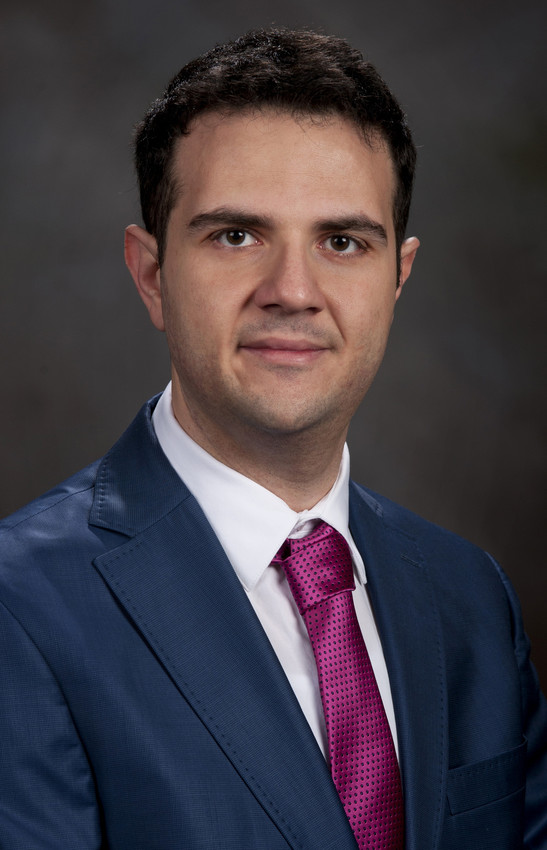}}]{Walid Saad } (S’07, M’10, SM’15, F’19) received his Ph.D. degree from the University of Oslo in 2010. Currently, he is a professor in the Department of Electrical and Computer Engineering at Virginia Tech. His research interests include wireless networks, machine learning, game theory, cybersecurity, unmanned aerial vehicles, and cyber-physical systems. He is the author/co-author of eight conference best paper awards and of the 2015 IEEE ComSoc Fred W. Ellersick Prize.
\end{IEEEbiography}

\begin{IEEEbiography}[{\includegraphics[width=1in,height=1.25in,clip,keepaspectratio]{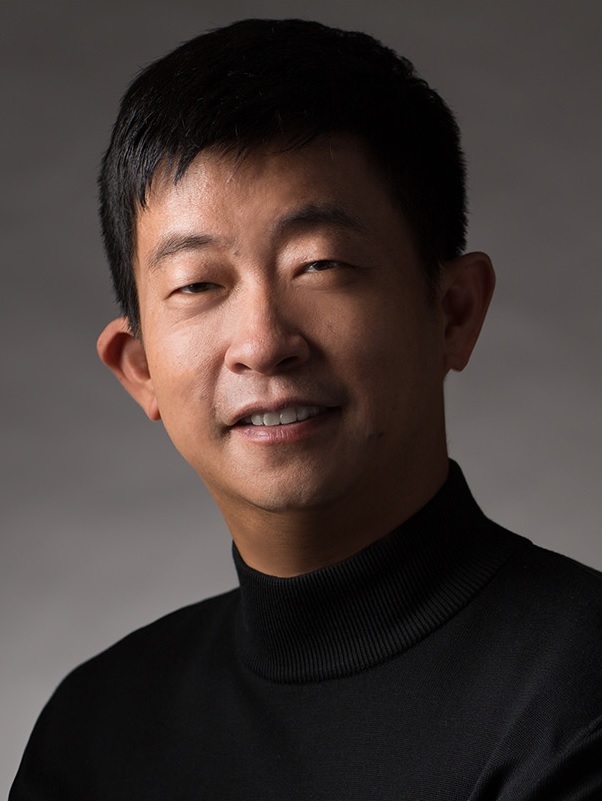}}]{Zhu Han}(S’01, M’04, SM’09, F’14) received his Ph.D. degree in electrical and computer engineering from the University of Maryland, College Park. Currently, he is a professor in the Electrical and Computer Engineering Department as well as in the Computer Science Department at the University of Houston, Texas. Dr. Han is an AAAS fellow since 2019. Dr. Han is 1\% highly cited researcher since 2017 according to Web of Science.

\end{IEEEbiography}

\begin{IEEEbiography}[{\includegraphics[width=1in,height=1.25in,clip,keepaspectratio]{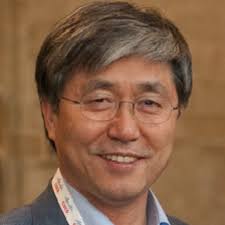}}]{Choong Seon Hong} (S’95-M’97-SM’11) is working as a professor with the Department of Computer Science and Engineering, Kyung Hee University. His research interests include future Internet, ad hoc networks, network management, and network security. He is currently an Associate Editor of the IEEE Transactions on Network and Service Management, International Journal of Network Management, and Journal of Communications and Networks and an Associate Technical Editor of the IEEE Communications Magazine.
\end{IEEEbiography}

\end{document}